\documentclass[aps,prb,twocolumn,showpacs]{revtex4}

\usepackage{graphicx}
\DeclareGraphicsExtensions{.png,.jpg,.eps}
\usepackage{xcolor}

\begin{document}

\title{Noncollinear magnetic phases and   edge states  in graphene quantum Hall bars}

\author{J. L. Lado, J. Fern\'andez-Rossier}
\affiliation{  International Iberian Nanotechnology Laboratory (INL),
Av. Mestre Jos\'e Veiga, 4715-330 Braga, Portugal
}

\date{\today} 

\begin{abstract} 

Application of a perpendicular magnetic field to charge neutral graphene is expected to result in a variety of broken symmetry phases, including  antiferromagnetic , canted and ferromagnetic.  All these phases 
open a gap in bulk but  have very different edge states and noncollinear spin order,  recently confirmed experimentally.
 Here we provide an integrated description of  both edge and bulk for the various magnetic phases of graphene Hall bars   making use of a    noncollinear mean field Hubbard model. 
 Our calculations   show that, at the edges,  the three types of magnetic order are either enhanced (zigzag) or suppressed (armchair). Interestingly,  we find that preformed local moments in zigzag edges interact with the quantum Spin Hall like edge  states of the ferromagnetic phase  and can induce back-scattering.

\end{abstract}
\maketitle

\section{Introduction}

\def \theorypapers {qh-inter,dimer,ferro-qh,qh-long-cou,zou,review-int,ex-gap,brey2006,herbut2,jeil-macdonald,macdonald-nomura,Alicea,Kharitonov,kharitonov2,bitan14}

The remarkably perfect\cite{VK} quantization of conductance in  quantum Hall systems, that provides our standard of resistance  for $h/e^2$ ,   arises from the  combined opening of a gap in the bulk of the sample, topologically different from vacuum, that implies the  existence of chiral edge states for which back-scattering is impossible\cite{Laughlin81,Halperin82}.  In graphene,  quantum Hall effect \cite{Novoselov05,Kim05} also shows perfect quantization, but it has its own peculiarities\cite{katsnelson,Gusynin05},   as a result of the relativistic-like 
nature of the graphene electron dispersion,  with two  non-equivalent valleys with Dirac-like bands. 
In a honeycomb lattice,   each valley hosts
two  sets of unevenly spaced Landau levels (LL), for electrons $n>0$ and holes $n<0$, plus a special
 $n=0$ LL,   at the Dirac energy.   Whereas all graphene LL have 
 fourfold degeneracy, coming from the spin and valley,  the $n=0$ is different from the rest: for a given spin and a given edge, it   has only one electron-like and hole like dispersive edge states.
 
    It was found\cite{levitov} early on that perturbations could  open a gap in the $n=0$ LL in two fundamentally different ways, either by splitting the levels according to their spin, or to the valley (which is completely correlated to the sub lattice for the $n=0$ LL),  resulting in  very different edge states.  Whereas breaking the sub lattice symmetry would give gaped edge states (leftmost panel of figure 1d), the spin-polarized state would have counter propagating edge states inside the gap (rightmost panel of figure 1d). In this sense,  the spectrum of spin-polarized graphene  in the quantum Hall regime, would be identical to the acclaimed  quantized Spin Hall phase proposed by Kane and Mele\cite{kane-mele}. 
  
  \begin{figure}
 \centering
                \includegraphics[width=.5\textwidth]{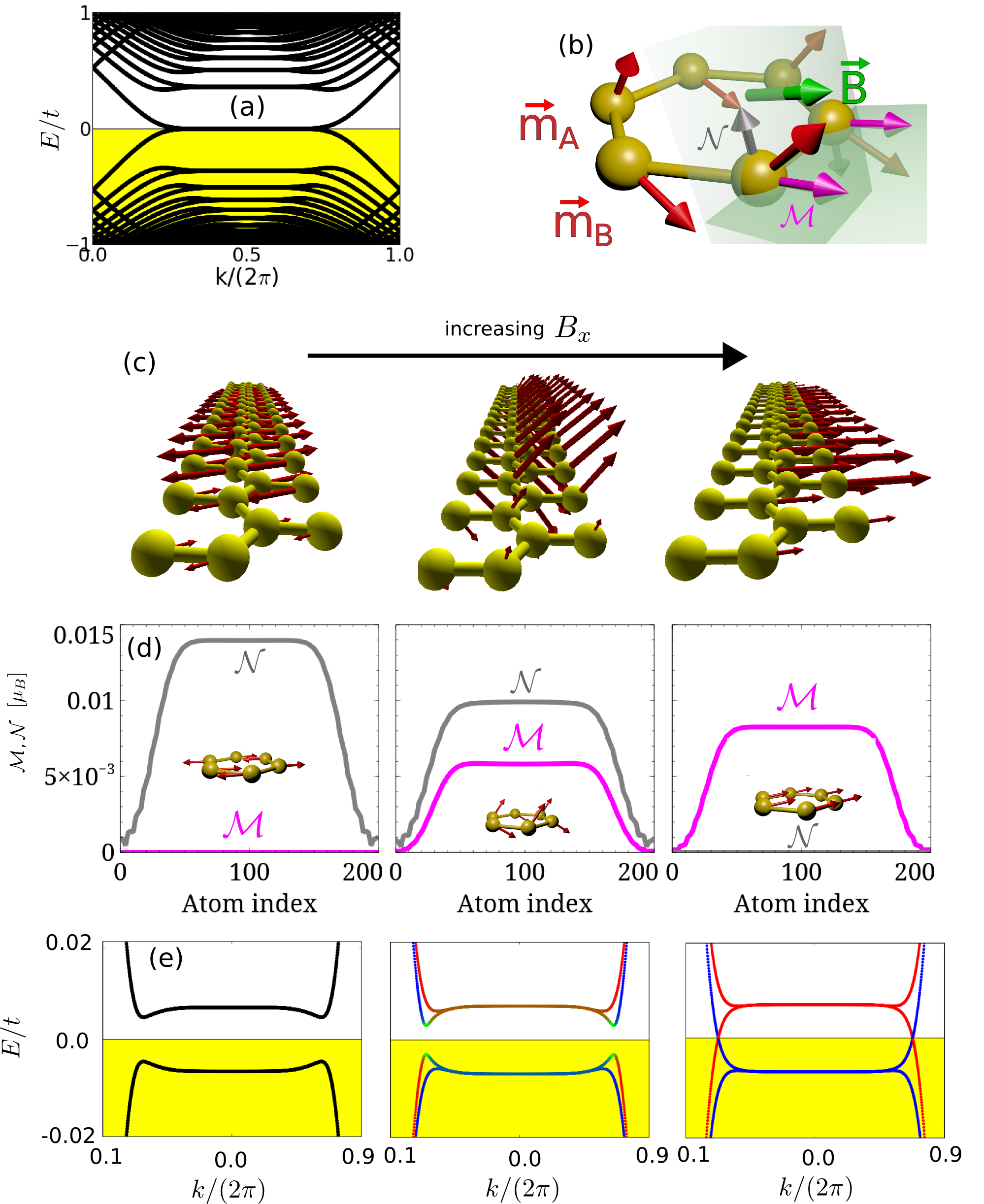}

\caption{(a) Non interacting spectrum of a quantum Hall armchair ribbon,
and (b) scheme of the magnetism developed when interactions, off-plane and
in-plane fields are
present. (c) Magnetism in the ribbon,
(d) magnetic order parameters and (e) band structures as the
in-plane field increases, showing a pure edge insulator to metal transition.}
\label{f1}
\end{figure}

A generic feature of quantum Hall systems, is that when the Fermi energy lies at the middle of a  Landau level,  interactions can open a gap and break the spin symmetry. In the case of spin-degenerate LL, 
this   leads to the  quantum Hall ferromagnetism.   In the case of the $n=0$ graphene quartet, several types of electronic order have been studied\cite{\theorypapers}. Three obvious candidates
come to mind:   a charge density wave (CDW), a spin density wave resulting in antiferromagnetic (AF) order, both  breaking valley
symmetry,  and ferromagnetic order (FM).  Only the latter is expected to have gap-less spin-filtered edge states
\cite{levitov,Kharitonov}. 

Different   experiments have provided  strong evidence of interaction driven band gap opening in the $n=0$ quartet in graphene  at half filling\cite{Kim2006,Kim2007,Strocio2010}. In a  recent experimental breakthrough\cite{young2014},   the combined  application of an in-plane $B_x$ and off-plane $B_z$ fields,   have made it possible
to observe the controlled  transition between different 
electronically ordered  phases at half filling.   Thus, as $B_x$ is ramped up, the system goes from 
 an insulating phase, most
likely AF,  to a phase with thermally activated edge transport, presumably
a canted AF (CAF) phase with gapped edge states, and at higher in-plane
field to a phase with $G$ slightly below $2G_0$ (where $G_0=e^2/h$) , as expected from the
FM phase with quantum spin Hall like spectrum. Upon gating, all these phases merge
into a phase with $G=G_0$.

These recent experimental results\cite{young2014}  highlight the interplay between noncollinear  bulk electronic  order and the emergence of spin-filtered  edge states that, in contrast with the $n\neq 0$ states that are topologically protected,  are the consequence of an interaction driven electronic phase transition in bulk. 
This  motivates the  interacting theory  presented here, that describes on equal footing the noncollinear spin order of  both bulk and edge states,  going beyond previous theory work \cite{\theorypapers}.
In particular,  a previously overlooked but important aspect of this problem is the fact that  the magnetic order is different at the edges and bulk.  Our noncollinear  mean field Hubbard model  calculations show that zigzag and armchair edges enhance and suppress, respectively, the magnetic order associated to the bulk Landau levels.  In the case of the recently observed Quantum Spin Hall like phase, the coexistence of spin filtered edge states with local magnetic moments at zigzag terminations is likely to play a role in the observation\cite{young2014} of a conductance smaller than $2G_0$ that is expected from these states in the absence of spin flip interactions\cite{levitov}. 

The manuscript is organized as follows. In Section II we present the
tight binding model and mean field approach used to model the Quantum Hall bars.
In section III we present the results for the edge and bulk electronic
properties at half filling, as well as for the $\nu=1$ phase.
Finally in Section IV we summarize our conclusions.

\section{Model}

We model graphene quantum Hall bars with a Hubbard model for a honeycomb lattice stripe: 
\begin{equation}
{\cal H}= {\cal H}_0(B_z) + g\mu_B \vec{B}\cdot\vec{S} + U \sum_i n_{i,\uparrow}n_{i,\downarrow}
\end{equation}
The first term describes electrons in a honeycomb lattice and the  effect of the perpendicular magnetic field $B_z$ on the orbital motion is included by means of the standard Peierls substitution in the tight binding model.  The second term is the Zeeman coupling and the third is the Hubbard term,  where $n_{i\sigma}=c^{\dagger}_{i\sigma}c_{i\sigma}$ is the occupation number for spin $\sigma$ at site $i$. 
Our gauge choice preserves translational invariance along the transport direction, so that $k$ is a good quantum number.
The    spectrum of ${\cal H}_0(B_z)$ is shown in figure 1a for a stripe
with armchair terminations,  and features both the bulk Landau levels and the dispersive edge states.

The effect of interactions is treated at the 
unrestricted Hartree-Fock approximation with a 
variational wave function 
$|\Omega\rangle=\prod_i \left(u_i c_{i\uparrow}^{\dagger} + v_i c_{i\downarrow}^{\dagger}\right) |0\rangle$. This naturally leads to write the Hubbard part of the Hamiltonian as a one-body  mean field Hamiltonian:
\begin{equation}
H^{MF}=H_0+g\mu_B\vec{B}\cdot\vec{S}+H_H+H_F+E_{DC}
\end{equation}
where $H_{H}=U\sum_{\sigma} n_{i\sigma} \langle n_{i\overline{\sigma}}\rangle$ is the Hartree term, 
$H_F=-U\sum_{\sigma} c^{\dagger}_{i\sigma} c_{i\overline{\sigma}}  
\langle  c^{\dagger}_{i\overline{\sigma}} c_{i\sigma}  \rangle$ is the Fock term, 
$E_{DC}=-U[\langle n_{i\uparrow} \rangle \langle n_{i\downarrow}\rangle-\langle
c^\dagger_{i\uparrow} c_{i\downarrow}\rangle \langle c^\dagger_{i\downarrow} c_{i\uparrow}\rangle]
$
is a constant,  $\overline{\sigma}=-\sigma$ and $\langle O \rangle=\langle \Omega|O|\Omega\rangle$.  The variational coefficients $v_i$ and $u_i$ are determined by iteration, starting from a trial solution, until a self-consistent solution is found. Solutions for bulk graphene, ignoring boundaries, can be found analytically\cite{herbut2} and are consistent with our numerical results. For strips,  a numerical implementation of this procedure yields, in general, 
 solutions with noncollinear magnetization whose magnitude and orientation vary from bulk to edge.   Since a numerical calculation of the actual stripes,  with one micron width, is beyond reach of  our computational resources, we consider
narrower stripes with  $W=$10 nm, with larger $B_z$, so that 
the magnetic length $\ell_B=\sqrt{\frac{\hbar}{eB_z}}$
 that controls inter edge coupling is still much smaller than W.

\begin{figure}
 \centering
                \includegraphics[width=.5\textwidth]{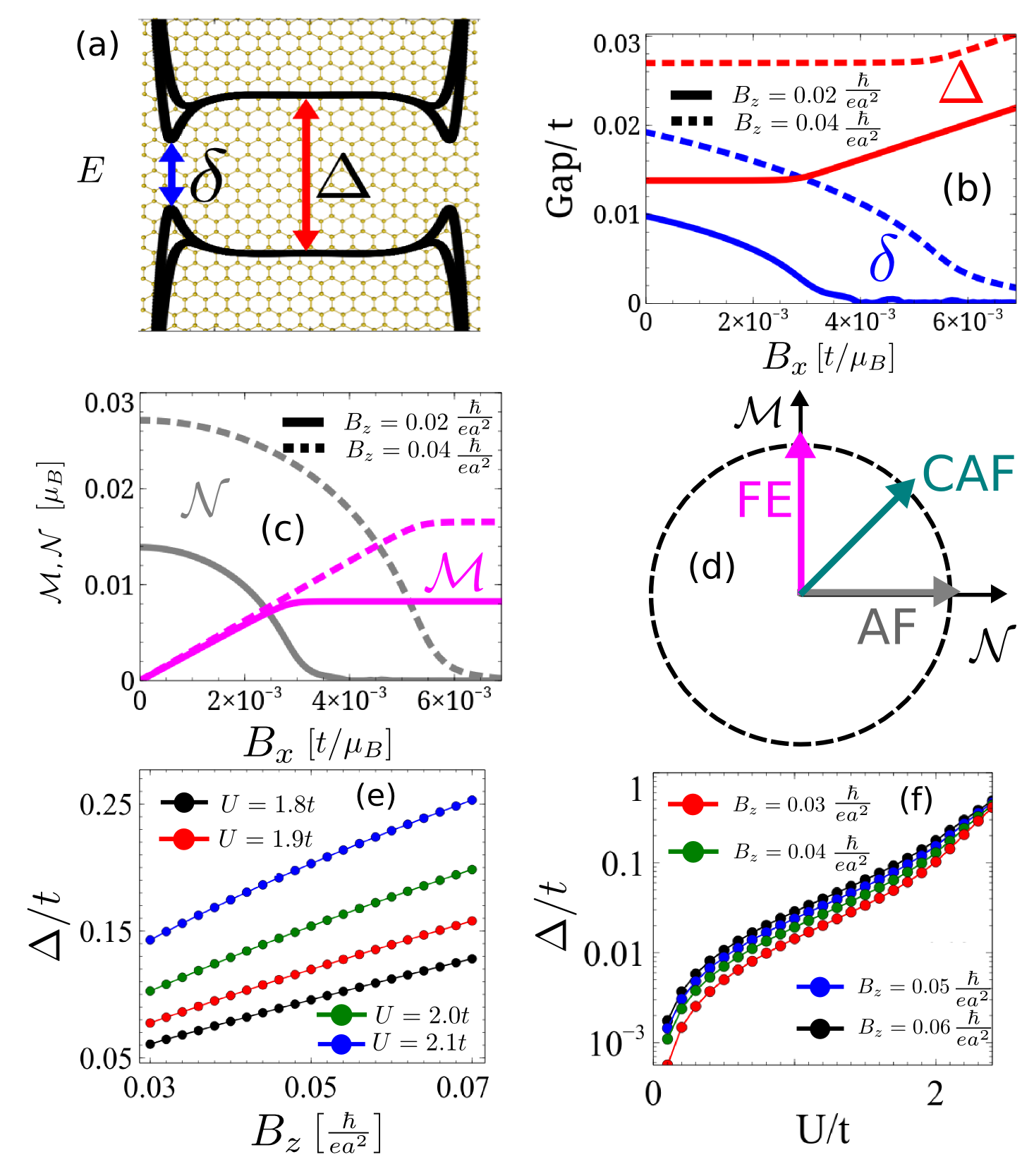}

\caption{
(a) Scheme
of the two different gaps (edge and bulk) observed in the band structure
for an armchair ribbon,
and (b) their
evolution with an increasing in-plane field.
(c) Bulk magnetic order parameters obtained in the calculation,
and (d) scheme representing
the phase transition in the order parameter space.
(e) Dependence of the AF gap in absence of in-plane field in
a quantum Hall armchair bar as a function of the off-plane field,
and (f) the electron-electron interaction.
}
\label{f2}
\end{figure}

\begin{figure*}[hbt]
 \centering
                \includegraphics[width=\textwidth]{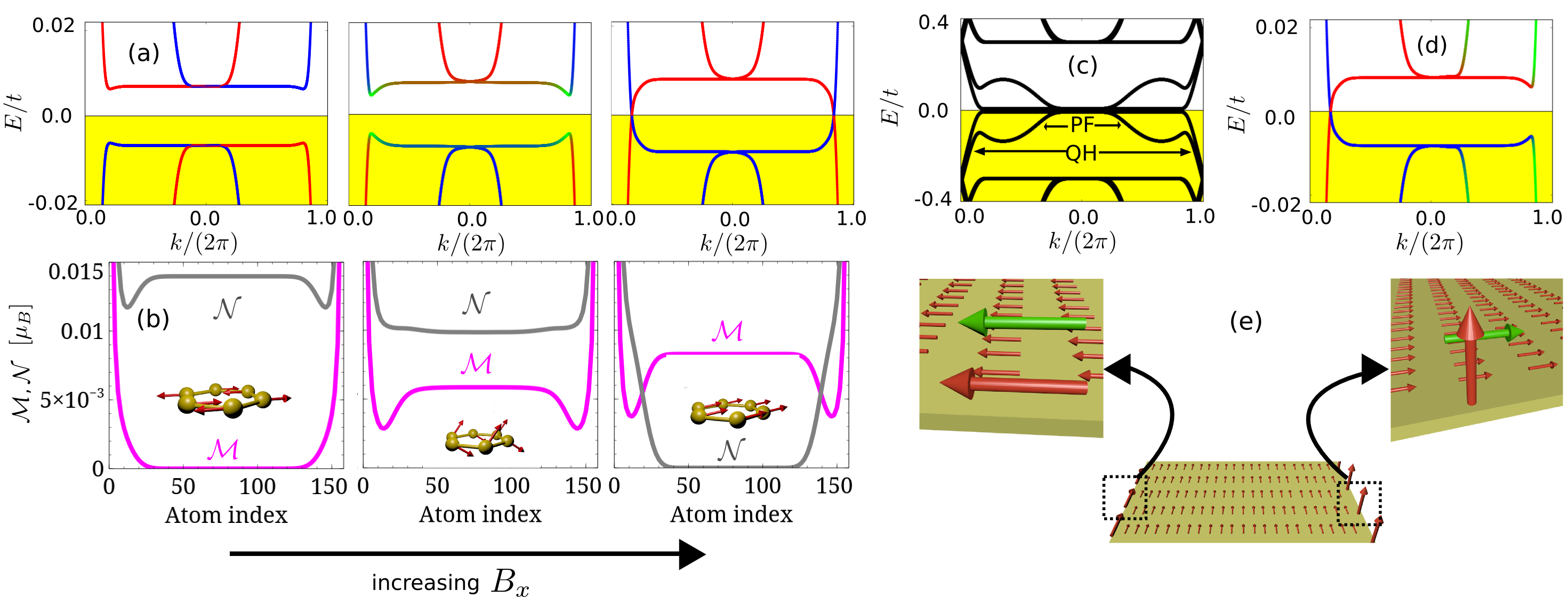}

\caption{
(a) Energy bands for the AF, CAF and FM in a zigzag ribbon and
(b) magnetic order parameters across the width. (c) Band structure of a zigzag
ribbon marking the two different kinds of edge states. (d) Band structure
and (e) scheme of a zigzag ribbon in the ferromagnetic regime with an excited
magnetic moment on one edge.
}
\label{f3}
\end{figure*}

The magnetic order of a given self-consistent solution is completely characterized by the average spin moment in every atom of the ribbon unit cell, $\vec{m}_i = \langle\Omega|\vec{S}_i|\Omega\rangle$. It is convenient to introduce to two fields that measure the degree of  ferromagnetic (FM) and antiferromagnetic (AF) order of a given solution. For each pair of adjacent atoms, $A$ and $B$, we define:
\begin{equation}
\vec{M}=\frac{\vec{m}_A+\vec{m}_B}{2}, \quad \,\,\,
\vec{N}=\frac{\vec{m}_A-\vec{m}_B}{2}
\end{equation}

\section{Results}

At half filling we find that $|\vec{m}_A|=|\vec{m}_B|$
which implies that $\vec{N}$ and $\vec{M}$ are orthogonal. We also find that  $\vec{N}$ and $\vec{M}$   points always perpendicular and parallel, respectively,  to the applied magnetic field, in order to  minimize the Zeeman energy.
The mean field Hamiltonian  is invariant to rotations of $\vec{N}$ in the plane perpendicular to the applied field.  Therefore, at half filling  is enough to refer to ${\cal N}=|\vec{N}|$ and ${\cal M}=|\vec{M}|$

\subsection{Bulk properties at half filling}

Both the evolution of the magnetic order parameters  from edge to edge [Figs. 1(d) and 3(b)], as well 
as the  band dispersion [Figs. 1(e) and 3(a)], make it clear
that edges and bulk are very different.  We first discuss the calculated properties
of  the bulk region, which are in line with previous theory work\cite{herbut2,Kharitonov}. At half filling we find three different phases,  depending on the value of $B_x$.  In the limits $B_x=0$ and $B_x\rightarrow\infty$, the  Hubbard  interaction yields bulk  in-plane antiferromagnetic order (${\cal M}=0$) and in-plane ferromagnetic order (${\cal N}=0$), respectively.
  As $B_x$ is ramped between these two extremes, both the ${\cal N}$ and ${\cal M}$ components survive, in the so called canted AF (CAF) phase\cite{Kharitonov,herbut2}. 
  
  The three magnetically ordered phases, FM, AF or CAF,  open  a bulk gap $\Delta$
  in the $n=0$ quartet. 
In Fig. 2(b) and 2(c) we show how the magnitude of the bulk gap $\Delta$ remains initially constant as a function of $B_x$, whereas the ${\cal M}$ increases linearly and ${\cal N}$ is depleted according to a $\sqrt{1-\left(\frac{B_x}{B_0}\right)^2} $ law.  Since ${\cal M}$ scales
linearly with $B_x$ this results shows that $\Delta \propto \sqrt{{\cal M}^2+{\lambda^2\cal N}^2}$,  and $B_x$ is actually driving
a rotation of the $({\cal N},{\cal M})\propto (cos\theta,\lambda sin\theta)$
vector\cite{herbut2}.
Thus, the FM, AF and CAF phases can be interpreted as different realizations of a common multidimensional order parameter, rather than phases with different order parameters\cite{herbut2,Kharitonov}.

An important test for the model is the dependence of the
bulk gap $\Delta$  on the off-plane magnetic field $B_z$. In the experiments, a roughly
linear dependence\cite{linear-scaling} $\Delta\propto B_z$ was found, in contrast
with the expected\cite{Alicea,macdonald-nomura}  from  the HF theory for long-range Coulomb interaction, in
which $\Delta\simeq \frac{e^2}{\ell_B}\propto\sqrt{B}$.

Within the mean field Hubbard model, the origin of
the linear scaling is the following. First, the gap scales
linearly with the atomic magnetic moment. Secondly, the magnetic moment
scales linearly with the number of electrons in the zero Landau level,
which are the ones that are unpaired.  This number of electrons
is given by the ratio of the area of sample and
the square of the magnetic length
$A_{sample}/l_B^2$.
This ratio is proportional to $B_z$, yielding the linear scaling of the gap.

The magnitude of the mean field gap depends strongly on $U$. In order to account for the experimentally observed  magnitude of $\Delta/(eBa^2/\hbar) = 40 eV$, we would need to assume  $U/t\simeq2-2.5 $, within the limits considered in the literature\cite{value-u}.  However,  in order to calculated $\Delta$ it is probably more realistic to  assume a  smaller value for $U$  and to include the effect of long range Coulomb interaction as well \cite{Alicea,macdonald-nomura}.

\subsection{Edge properties at half filling}
   
  We now discuss the electronic properties of the edges.  We  consider both zigzag and armchair terminations. 
 In both cases ${\cal N}$ and ${\cal M}$ are  modulated as the edge is approached, but in a different way:  they are  depleted  at the armchair edges [Fig. 1(c,d)]
and  enhanced at the zigzag terminations [Fig. 3(b)].
 We start the discussion with  the evolution of the edge states as a function of $B_x$ for the simpler case of armchair edges. 
  As $B_x$ is ramped up, ${\cal N}$ is depleted in bulk, so it does
the edge gap, $\delta$ [Fig. 2(b)]. Thus, in the AF phase, with $\delta\simeq \Delta>> k_BT$ edges are insulating, but in the CAF phase,  when ${\cal N}$  and $\delta$ are  close to zero, thermally activated edge transport is possible, as reported experimentally\cite{young2014}.   
In the ferromagnetic phase, with ${\cal N}=0$,  the edge gap closes, $\delta=0$, and our calculated spectrum  is identical to that of a Quantum Spin Hall insulator\cite{kane-mele}, with a finite gap $\Delta$ in the  bulk spectrum and spin-polarized counter
propagating gap-less edge states.

The existence of the  quantum spin Hall like spectrum  in the FM phase is  true both for zigzag and armchair  terminations,   confirming the prediction  based in a model that ignored the modulation of the order parameter at the edge\cite{levitov}.  However,   in the zigzag edges the  enhancement of the magnetic moment at the edges has non-trivial and important consequences. 
Two types of edge states exist at the zigzag terminations in the Quantum Hall regime: the topologically protected current carrying states,  present in any Quantum Hall bar (QH), and the preformed (PF)  non-dispersive mid gap edge states\cite{Nakada96}, present already at $B_z=0$ , that host magnetic moments when Hubbard interactions are turned on\cite{Fujita96,JFR08}. These two types of edge states, QH and PF, are marked in Fig. 3(c).  Inspection of their wave function confirms that the PF edge states are mostly localized in last atomic row of the stripe, in contrast with the QH states, that are extended over a distance of order  $\ell_B$\cite{Brey06} .
The enhancement of the magnetic order at the zigzag edges comes from the
interaction driven ferromagnetic order associated to the PF edge states. The calculations yield 
edge magnetic moments lying parallel to the applied magnetic field, i.e., mostly in-plane.

   An important question is to which point the edge magnetic moments associated to the PF states are
  coupled to the QH edge states.
 To address this question we perform a mean field calculation for the FM phase (with $B_x>>B_z$) where we constraint the magnetic moment of one of the edges  to lie perpendicular to the plane [Fig. 3(e)]. The resulting self-consistent solution still has in-plane magnetization for bulk and for the free edge. 
  The calculated energy bands are shown in Fig. 3(d). It is apparent that at the edge where the PF magnetic moments are forced to lie off plane,   a gap opens in the QH edge states.
  
  Our  calculation clearly shows that spin-filtered QH edge states are sensitive  to the spin orientation of the PF edge moments.   This provides a natural scenario to account for the 
 experimentally observed value \cite{young2014}  of the zero  bias conductance $G=1.8 G_0$,  rather than $G=2G_0$, the conductance expected if no spin-flip interactions occur.   In a micron size flake there will be several patches with zigzag terminations and PF edge states. There, spin fluctuations of the edge moments will induce spin mixing and back-scattering of the spin-filtered QH edge states.   
  A second mechanism  that 
 would induce spin back-scattering  combines  spin-orbit coupling and disorder\cite{inpreparation}.

\begin{figure}[hbt]
 \centering
                \includegraphics[width=.5\textwidth]{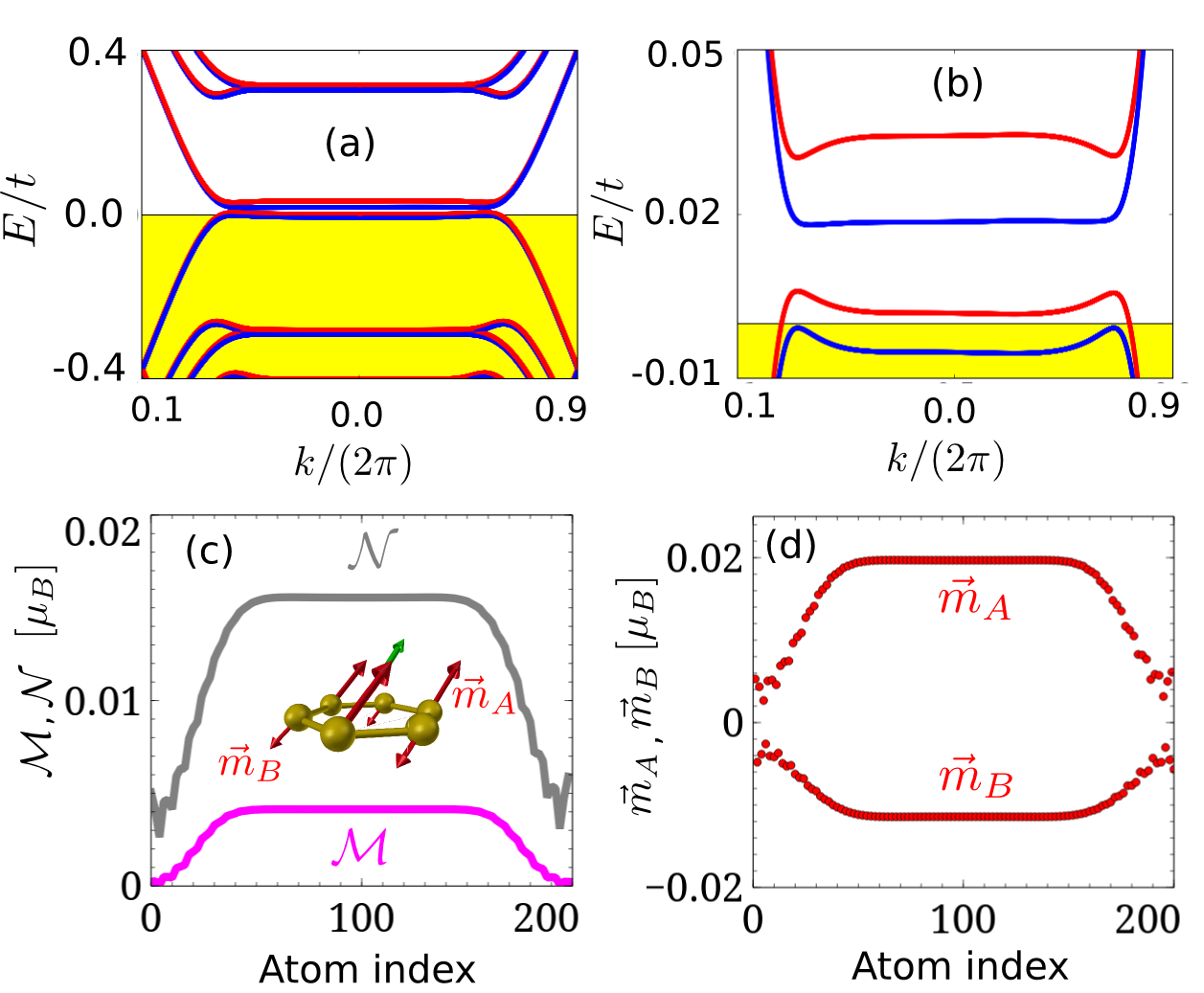}

\caption{ Quarter filling phase. 
(a,b) Energy bands,  (c) magnetic order parameters, ${\cal N}$ and ${\cal M}$ as a function of position and (d) sublattice resolved magnetic moments. } 
\label{f4}
\end{figure}

\subsection{ Ferrimagnetic $\nu=1$ phase away from half filling}

As a final test for the model,  we now discuss our results [Fig. 4] for the system away from half-filling. 
Experiments\cite{young2014}  indicate that,  upon gating up to quarter filling 
 the three magnetic phases merge into a unique phase with edge
conductance $G=G_0$ which means that both spin and valley degeneracy have to be broken. 
 Our calculations,
show that at quarter filling [Figs. (a) and (b)],
the system develops a ferrimagnetic
phase [Fig. 4(c)]. Unlike  the case of half-filling, we
have now $|\vec{m}_A|\neq|\vec{m}_B|$ and the magnetic moments of both
sub lattices are parallel to the total applied field.  The different magnitude of the sub lattice magnetizations $m_A$ and $m_B$, shown in Fig. 4(d), is a clear indication of the valley symmetry breaking.   

The occupation of a unique LL, possible due to the valley and spin symmetry breaking,  automatically
implies that a single  spin-polarized dispersive edge state,  accounting for the observation\cite{young2014} of the $G=G_0$ plateau.
Importantly,  both the bulk magnetization and
the number of edge channels  in this ferrimagnetic phase are
insensitive to the magnitude of the in plane
magnetic field, in agreement with the  experiment \cite{young2014}.   Moreover, the bulk gap
is found to increase with the in-plane field as observed in experiments.


\section{Summary}

We have presented a comprehensive study of magnetic order in graphene stripes in the Quantum Hall regime, mostly at half filling,   based on a noncollinear mean field Hubbard model that treats both edge and bulk on equal footing.   The interplay between bulk magnetism and the type of edge state, confirmed in   recent experimental results\cite{young2014},  motivates the present work where the modulation of the magnetic order at the edges is taken into account.    The model captures the main experimental observations including the linear scaling between the bulk gap $\Delta$ and the off-plane magnetic field $B_z$.  
Our calculations reveal the coupling between the quantum spin Hall like edge states and preformed local moments at zigzag edges, that provides a natural scenario for the spin back-scattering observed experimentally.  This scenario is similar to recent proposals\cite{Gloria12,Altshuler13} where spin-Hall edge states interact with magnetic impurities.  The last but not the least, at quarter filling the model predicts the existence of a previously overlooked ferrimagnetic phase with spin polarized edge states with $G=G_0$.

We thank J. W. Gonz\'alez for fruitful discussions.  We acknowledge financial support by Marie-Curie-ITN 607904-SPINOGRAPH.
 JFR acknowledges  financial support   by Generalitat Valenciana (ACOMP/2010/070), Prometeo. 

\end{document}